\documentclass[conference]{IEEEtran}

\usepackage[noadjust]{cite}
\usepackage{blindtext} 
\usepackage{amsmath} 
\usepackage{amsfonts} 
\usepackage{graphicx,epstopdf} 
\usepackage{tabulary}
\usepackage{xcolor} 

\ifCLASSOPTIONcompsoc 
\usepackage[caption=false,font=normalsize,labelfont=sf,textfont=sf]{subfig}
\else
\usepackage[caption=false,font=footnotesize]{subfig}
\fi

\newcommand{\tsup}{\textsuperscript}
\newcommand{\tsub}{\textsubscript}
\DeclareMathOperator*{\argmax}{arg\,max} 

\newcommand{\rom}[1]{\uppercase\expandafter{\romannumeral #1\relax}}

\begin{document}
	\bstctlcite{BSTcontrol} 
	
	\title{RACH Optimization with Decision Tree Based Supervised Learning for Conditional Handover in 5G Beamformed Systems}
	
	\author{\IEEEauthorblockN{
			Umur Karabulut\IEEEauthorrefmark{1}\IEEEauthorrefmark{2}, 
			Ahmad Awada\IEEEauthorrefmark{1}, 
			Ingo Viering\IEEEauthorrefmark{3}, 
			Andre Noll Barreto\IEEEauthorrefmark{4} and 
			Gerhard P. Fettweis\IEEEauthorrefmark{2}}
		\IEEEauthorblockA{
			\IEEEauthorrefmark{1}Nokia Bell Labs, Munich, Germany, 
			\IEEEauthorrefmark{2}Vodafone Chair Mobile Communications Systems, Technische Universit\"at Dresden,\\
			\IEEEauthorrefmark{3}Nomor Research GmbH, Munich, Germany, 
			\IEEEauthorrefmark{4}Barkhausen Institut gGmbH, Dresden, Germany
		}
	}
	
	\maketitle
	
	\begin{abstract}
		Higher frequencies that are introduced in 5G networks cause rapid signal degradation and challenge user mobility. In recent studies, a conditional handover procedure has been adopted for 5G networks to enhance user mobility robustness.
		In this paper, mobility performance of the conditional handover is analysed for 5G mm-Wave systems with beamforming. In addition, a resource efficient random access procedure is proposed that increases the chance of contention-free random access during handover, which reduces signaling and interruption time. Moreover, simple, yet, effective decision tree based supervised learning method is proposed to minimize the handover failures that are caused by beam preparation phase of random access procedure. Results reveal the trade-off between contention free random access and handover failures. It is also shown that the optimum operation point of random access is achievable with proposed learning algorithm for conditional handover.
	\end{abstract}
	
	
	\section{Introduction}
		In cellular networks, demand for user data throughput will continue to increase dramatically \cite{Throughput}. The range of carrier frequency has been further expanded to mm-Wave frequencies in fifth generation (5G) cellular networks to meet the increasing demand of user data throughput. In addition, the number of base stations (BSs) with smaller coverage area is increased which improves frequency reuse and the total network capacity. Besides, higher carrier frequencies enable the deployment of many small-sized antennas that are used for directional signal transmission, resulting in beamforming gain. 
		
		Operating at higher carrier frequencies challenges user mobility due to steep and high diffraction loss which can lead to rapid signal degradation caused by obstacles \cite{HOsurvey}. Moreover, dense BS deployment increases the number of handovers which can cause frequent interruption of the user equipment (UE) connection, signaling overhead and latency \cite{HOsurvey}. 
		
		Baseline handover (BHO) procedure that is used in Long Term Evolution (LTE) is reused for 5G networks in the 3\tsup{rd} Generation Partnership Project (3GPP) release 15 \cite{38300,38331}. The time instant for triggering the handover in BHO is critical. This is because the signal of the serving cell should be good enough to receive the handover command and the signal of the target cell should be sufficient for access. This is more pronounced in mm-Wave frequencies due to the rapid signal degradations and dense BS deployment. 
		
		Conditional handover (CHO) is introduced in \cite{CHOTechRep} for New Radio (NR) 3GPP release 16 to increase the mobility robustness of the BHO. In CHO, the coupling between handover preparation and execution is resolved by introducing a conditional procedure, where handover is prepared early by the serving cell and access to the target cell is performed later when its radio link is sufficient. Furthermore, a contention-free random access (CFRA) procedure is defined in \cite{38331} where the target cell of the handover can allocate CFRA resources for the UE during the handover. Using CFRA instead of contention-based random access (CBRA) resources helps to avoid collision in random access, and consequently, mobility interruption and signaling overhead. On the other hand, persisting on CFRA resource usage during the handover could cause consecutive failure of random access which leads failure of handover process. 
		
		In this paper, a Resource Efficient (RE) Random Access Channel (RACH) procedure is proposed such that the utilization of CFRA resources is increased. Moreover, Beam-specific Enhanced Logging and Learning (BELL) approach is proposed which is a decision tree based supervised learning algorithm \cite{SupervisedLearning} to optimize the mobility performance of RACH procedure for CHO in beamformed systems. Besides, the mobility performance of CHO is analyzed first time in this paper for current 3GPP RACH procedure in comparison with RE RACH procedure and BHO.
		
		The paper is organized as follows. The UE measurements that are used for handover are presented along with BHO and CHO in Section~\ref{sec:Meas_HO}. The random access procedure that is defined in 3GPP is revisited and our RE RACH procedure is presented in Section \ref{sec:RACH}. The simulation scenario is explained in Section \ref{sec:SimScenario}. Simulation results are presented in Section \ref{sec:PerformanceEval} to show the performance of 3GPP and RE RACH procedures along with the BELL approach for CHO and BHO in 5G mm-Wave networks for different random access procedures. The paper is concluded in Section \ref{sec:Concl}.

	\section{UE Measurements and Handover Models}
	\label{sec:Meas_HO}
	
		In mobile networks, it is necessary to hand off the link of a UE between cells to sustain the user connection with the network. This handover is performed using UE received signal power measurements for serving and neighboring cells and by following a predefined handover procedure. In this section, baseline handover and conditional handover procedures are reviewed along with the relevant UE measurements for mobility.
				
		\subsection{UE Measurements in New Radio Beamforming System}
		A UE $ u $ in the network monitors the Reference Signal Received Power (RSRP) $ P_{c,b}^\textrm{RSRP}(n) $ (in dBm) at discrete time instant $ n $ for beams $ \forall b\in B $ of cell $ \forall c \in C$. The separation between the instants is given by $ \Delta t $ ms. The physical raw RSRP measurements are inadequate for handover decisions since those measurements fluctuate over time due to fast fading and measurement errors which would lead to instable handover decisions. To mitigate those channel impairments, UE applies a moving average Layer-1 (L1) filter and an infinite impulse response (IIR) Layer-3 (L3) filter to RSRP measurements sequentially. The implementation of L1 filtering is not specified in 3GPP standardization and it is UE specific, i.e., it can be performed either in linear or dB domain. The L1 filter output can be expressed as
		\begin{equation}\label{eq:L1filter}
		P_{c,b}^\textrm{L1}(m) = \frac{1}{N_\textrm{L1}}\sum_{\kappa = 0}^{N_\textrm{L1}-1}P_{c,b}^\textrm{RSRP}(m-\kappa),~ m=n\omega
		\end{equation}
		where $ \omega \in \mathbb{N}$ is the L1 measurement period normalized by time step duration $ \Delta t $ , and $ N_\textrm{L1} $ is the number of samples that are averaged in each L1 measurement period. For cell quality derivation of cell $ c $, set $ B^\textrm{str}_c $ of beams having measurements above threshold $ P_\textrm{thr} $ is determined as
		\begin{subequations}
			\begin{align}\label{eq:BeamConsol1}
				& B^\textrm{str}_c(m) = \{b~|~P_{c,b}^\textrm{L1}(m) > P_\textrm{thr}\}\\
				\intertext{subject to}
				&P_{c,b_i}^\textrm{L1}(m) > P_{c,b_j}^\textrm{L1}(m),~ \forall b_i \in B^\textrm{str}_c,~\forall b_j \in \{B\setminus B^\textrm{str}_c\},\\
				&|B^\textrm{str}_c| \leq N_\textrm{str}, ~N_\textrm{str}\in \mathbb{N^+}.
			\end{align}
		\end{subequations}
		Cardinality of the set is denoted by $ \vert\cdot\vert $ and $ N_\textrm{str} $ is the maximum number of beams that are accounted for cell quality derivation. L1 RSRP measurement of beams $ b\in B_\textrm{str,c} $ are averaged to derive L1 cell quality of cell $ c $ as
		\begin{equation}\label{eq:BeamConsol2}
		P_c^\textrm{L1}(m) = \frac{1}{\vert B^\textrm{str}_c(m)\vert}\sum_{b\in B^\textrm{str}_c(m)}P_{c,b}^\textrm{L1}(m). 
		\end{equation}
		If $ B^\textrm{str}_c(m) $ is empty, $ P_c^\textrm{L1}(m) $ is equal to highest $ P_{c,b}^\textrm{L1}(m) $.
		
		L1 cell quality is further smoothed by L3 filtering and L3 cell quality output is derived by the UE as
		\begin{equation}\label{eq:L3Cell}
		P_c^\textrm{L3}(m) = \alpha P_c^\textrm{L1}(m) + (1-\alpha)P_c^\textrm{L3}(m-\omega),
		\end{equation}
		where $\alpha = \left( \frac{1}{2}\right)^\frac{k}{4}$ is the forgetting factor that controls the impact of older measurements $P_c^\textrm{L3}(m-\omega)$ and $ k $ is the filter coefficient of the IIR filter \cite{38331}.
		
		Similarly, the L3 beam measurement $ P_{c,b}^\textrm{L3}(m) $ of each beam is evaluated by L3 filtering of L1 RSRP beam measurements as
		\begin{equation}\label{eq:L3Beam}
		P_{c,b}^\textrm{L3}(m) = \alpha' P_{c,b}^\textrm{L1}(m) + (1-\alpha')P_{c,b}^\textrm{L3}(m-\omega),
		\end{equation}
		where $ \alpha' $ can be configured separately from $ \alpha $.
		
\begin{figure*}[!htb]
		\centering
		\includegraphics*[width=0.9\textwidth]{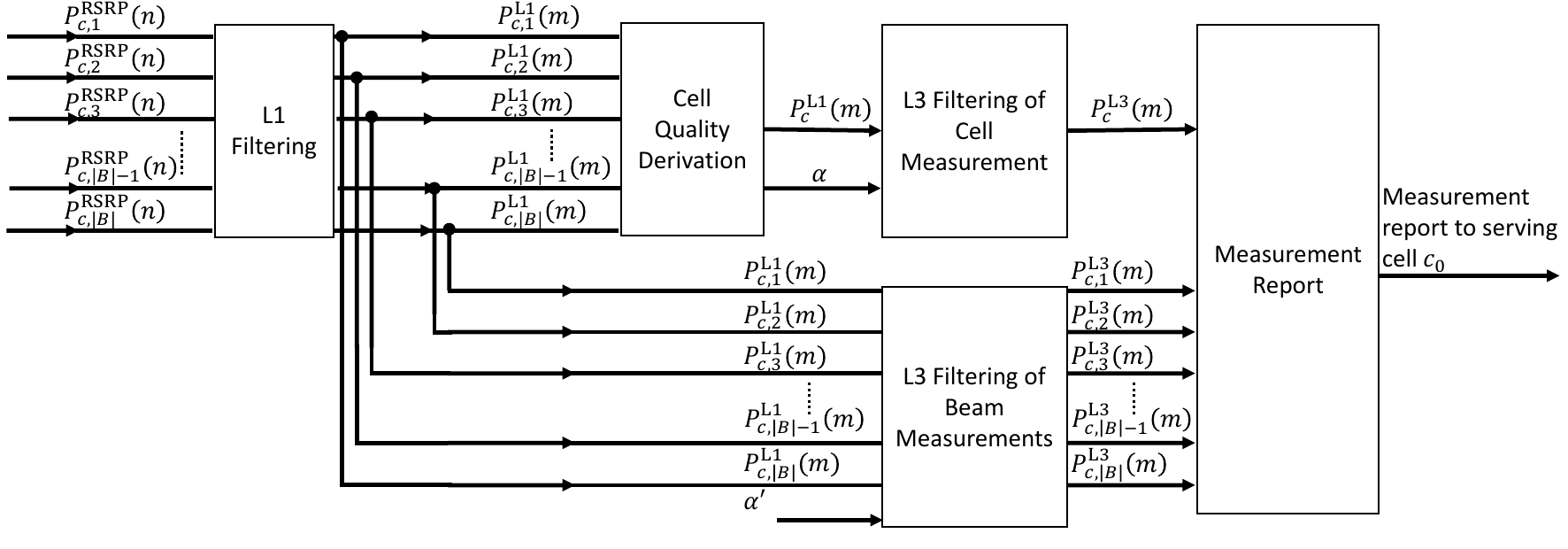}
		\caption{Diagram of L1 and L3 UE measurements which are derived from Reference Signal Received Power (RSRP) for beams of cell $ c $.}
		\label{fig:UEMeasurements}
		\vspace{-1\baselineskip}
		\end{figure*}

		L1 RSRP beam measurements $ P_{c,b}^\textrm{L1}(m) $, L3 cell quality measurements $ P_c^\textrm{L3}(m) $ and L3 beam measurements $ P_{c,b}^\textrm{L3}(m) $ that are used during the handover and the RACH procedure are illustrated in Figure~\ref{fig:UEMeasurements}.
		
		\subsection{Baseline Handover}
			
		L3 cell quality measurements $ P_{c}^\textrm{L3}(m) $ are used to assess the quality of the radio links between the UE and its serving and neighboring cells. To this end, UE reports the L3 cell quality measurements $P_{c}^\textrm{L3}(m) $ and beam measurements $ P_{c,b}^\textrm{L3}(m) $ to its serving cell $ c_0 $ if the following condition (A3)
		\begin{equation}
			\centering
			\label{eq:A3}
			P_{c_0}^\textrm{L3}(m) + o_{c_0,c}^\textrm{A\tsub3}< P_c^\textrm{L3}(m) ~~\textrm{for}~~ m_0-T_\textrm{TTT,A\tsub 3}<m<m_0,
		\end{equation}
		expires at time instant $ m = m_0 $ for any neighboring cell $ c\neq c_0 $. The cell-pair specific offset $ o_{c_0,c}^\textrm{A\tsub3} $ can be configured differently by serving cell $ c_0 $ for each neighboring cell $ c $ and time-to-trigger $ T_\textrm{TTT,A\tsub 3} $ is the observation period of condition (\ref{eq:A3}) before triggering measurement report. 
		
		After receiving  L3 cell quality measurements, the serving cell sends a handover request to a target cell $ c_T $, e.g., typically the strongest cell, along with the L3 beam measurements $ P_{c_T,b}^\textrm{L3}(m)$. Then, the target cell reserves CFRA resources (preambles) for beams $ b\in B^\textrm{prep}_{c_T} $ with the highest power based on reported $ P_{c,b}^\textrm{L3}(m)$. The target cell prepares the handover command including reserved CFRA resources and sends it to the serving cell as part of the preparation acknowledgment. After that, the serving cell sends the handover command to the UE. The command comprises the target cell configuration and CFRA preambles that are reserved by the target cell $ c_T $. After receiving the handover command, the UE detaches from the serving cell and initiates the random access towards the target cell. 
		
		In this handover scheme, the radio link between UE and serving cell  should be good enough to send the measurement report in the uplink and receive the handover command in the downlink. This is a necessary but not sufficient condition for completing the handover successfully. In addition, the radio link quality between the UE and the target cell should also be sufficient so that the signaling between UE and the target cell is sustained during the RACH procedure. In a typical system level mobility simulation, the link quality of the UE is assessed by Signal-to-Interference-Noise Ratio (SINR). Herein, the link quality conditions for successful handover between serving cell $ c_0 $ and target cell $c_T$ are expressed as
		\begin{subequations}
		\begin{align}
		\gamma_{c_0,b}(m_0) >& \gamma_{out}, \label{eq:ServingCond}\\
		\gamma_{c_0,b}(m_0+T_p) >& \gamma_{out},~~ \label{eq:ServingCond2}\\
		\gamma_{c_T,b}(m_0+T_p) >& \gamma_{out}  \label{eq:TargetCond},
		\end{align}
		\end{subequations}
		where $ \gamma_{c,b}(m) $ and $ \gamma_{c_0,b}(m) $, are the SINR of the links between UE and the beam $ b $ of target cell $ c_T $ and serving cell $ c_0 $, at time $ m $, respectively. $ m_0 $ is the time instant the measurement report is sent and $ T_p $ is the latency of handover preparation between serving and target cell. $ \gamma_{out} $ is the SINR threshold that is required for maintaining radio communication between UE and network (e.g. $ -8 $ dB). 
		
		As shown in (\ref{eq:ServingCond}), (\ref{eq:ServingCond2}) and (\ref{eq:TargetCond}), the time instant $ m_0 $ for triggering the measurement report is critical for the success of handover. When moving towards the coverage area of the target cell, delaying $ m_0 $ helps the conditions in (\ref{eq:ServingCond2}) and (\ref{eq:TargetCond}) to be fulfilled for serving cell $ c_0 $ and target cell $ c_T $, respectively, at the expense of having weaker $ \gamma_{c_0,b}(m_0) $ for serving cell risking the condition of (\ref{eq:ServingCond}), and vice-verse.
	
		\subsection{Conditional Handover}
		In conditional handover, the handover preparation and execution phases are de-coupled, which helps to receive the handover command safely from the serving cell and to access the target cell later when its radio link is sufficient.
		
		Similar to A3 condition (\ref{eq:A3}), an \textit{Add} condition is defined as,
		\begin{equation}
			\label{eq:Add}
			P_{c_0}^\textrm{L3}(m) + o_{c_0,c}^\textrm{add}< P_c^\textrm{L3}(m) ~~\textrm{for}~~ m_0-T_\textrm{TTT,add}<m<m_0,
		\end{equation}
		where $ o_{c_0,c}^\textrm{add} $ is defined as add offset. The UE sends the measurement report to serving cell $ c_0 $ at $ m = m_0 $ if the \textit{Add} condition is fulfilled for $ T_\textrm{TTT,add} $ seconds. Then, the serving cell $ c_0 $  sends the handover request to the target cell $ c_T $ for the given UE. The preparation of the handover is performed as in the baseline handover, where the target cell reserves CFRA RACH resources for the UE and sends the handover command to the UE via the serving cell. Unlike baseline handover, the UE does not detach from the serving cell immediately and initiates the RACH process towards the target cell when handover command is received. Instead, the UE continues measuring received signals from neighboring cells and initiates the random access when the \textit{Execution} condition expires at time instant $ m_1 $, after $ T_\textrm{TTT,exe} $ which is defined as,
		\begin{equation}
		\label{eq:Exec}
			P_{c_0}^\textrm{L3}(m) + o_{c_0,c_T}^\textrm{exe}< P_{c_T}^\textrm{L3}(m) ~~\textrm{for}~~ m_1-T_\textrm{TTT,exe}<m<m_1.
		\end{equation}
		The \textit{Execution} condition offset $ o_{c_0,{c_T}}^\textrm{exe} $ is configured by the serving cell and forwarded to the UE in the handover command along with CFRA resources reserved by the target cell. 
		
		Smaller $ o_{c_0,{c_T}}^\textrm{add} $ values lead to early preparation of the target cell and reservation of the RACH preambles which ensures that the UE sends the measurement report and receives the handover command (see (\ref{eq:ServingCond}) and (\ref{eq:ServingCond2})). Besides, unlike baseline handover, lower $ o_{c_0,{c_T}}^\textrm{add} $ does not lead to any early RACH attempt of the UE towards the target cell since the random access is initiated only if the \textit{Execution} condition is fulfilled. Higher \textit{Execution} condition offset $ o_{c_0,{c_T}}^\textrm{exe} $ values cause the UE to perform random access late enough such that it is more likely that the $ \gamma_{c_T,b}(m) $ is above $ \gamma_{out} $, see(\ref{eq:TargetCond}).

	\section{RACH Procedure in New Radio Multi-Beam System}
	\label{sec:RACH}
	In this section, the basics of random access are reviewed. Then, the 3GPP RACH procedure of NR \cite{38331} is described and our proposed RACH procedure is introduced. 
	
	\subsection{Contention-free and Contention-based Random Access}
	
	Random access is the first signaling performed by a UE for establishing the synchronization with a cell. The UE initiates the random access by sending a RACH preamble to the target cell. However, it is possible that multiple UEs use the same preamble during the random access towards the same reception beam of a target cell. In this case, RACH collision occurs which is then resolved by additional signaling and delay for completing random access. This type of random access where a UE selects one preamble out of set that is common for all UEs is called CBRA. 
	
	In handover, the collision risk can be avoided by assigning dedicated preambles to each UE to be used towards a prepared beam $ b\in B^\textrm{prep}_{c_T} $ of the target cell $ c_T $. The network identifies the UE signal without further signaling and delay if the UE accesses the prepared beam using the dedicated preamble. This type of random access is called CFRA. In \cite{38331}, $ B^\textrm{prep}_c $ is defined as
	\begin{subequations}
		\begin{align}\label{eq:PrepBeamSet}
			&B^\textrm{prep}_c = \{b |P_{b,c}^\textrm{L3}(m_0) \geq  P_{c,b_i}^\textrm{L3}(m_0),~b\neq b_i,~b,b_i\in B\}\\
			\intertext{subject to}
			&|B^\textrm{prep}_c| = N_\textrm{B}
		\end{align}
	\end{subequations}
where $ N_\textrm{B} $ is the number of beams that are prepared for random access.

	\subsection{Access Beam and Preamble Selection}
	\label{subsec:AccessPreamble}
	During handover, accessing the target cell by using a dedicated CFRA preamble is preferable due to lower latency and signaling requirements than CBRA. Although a set of beams $ b\in B^\textrm{prep}_{c_T} $ of the target cell $ c_T $ with the strongest L3 beam quality measurements $ P_{b,{c_T}}^\textrm{L3} $ can be prepared with CFRA resources, measurements of those beams may vary between the preparation time instant $ m = m_0 $ and access time $ m = m_1  $ due to the de-coupling between the phases. Variation of beam measurements is more significant in conditional handover compared to baseline handover. This is because, in baseline handover, the elapsed time between the preparation and access phases is given by $ T_p $ in (\ref{eq:TargetCond}). However, in conditional handover, this time is longer than $ T_p $ since the UE waits $ T_o $ period of time until the \textit{Execution} condition in (\ref{eq:Exec}) is fulfilled after receiving the handover command. In CHO, $ T_0 $ can be much larger than $ T_p $.
	
	\begin{figure}[!htb]
	\centering
	\includegraphics[width=0.8\columnwidth]{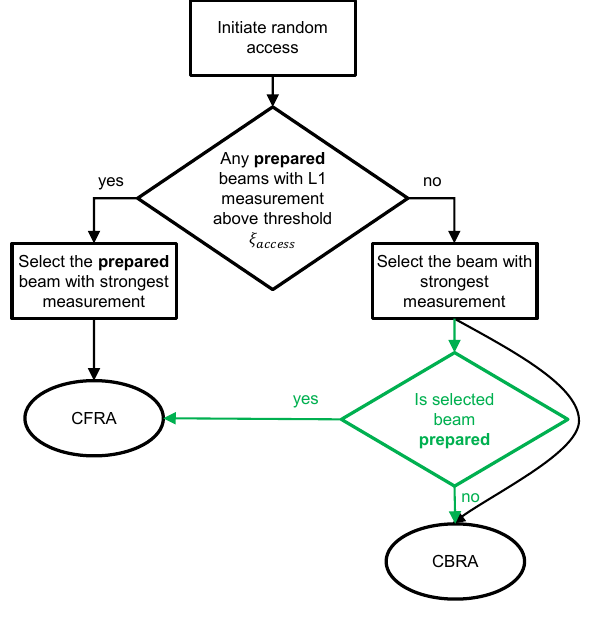}
	\caption{Random access flow diagram. The diagram shown in black is defined in 3GPP standardization and the green block is the proposed enhancement for the random access procedure.}
	\label{fig:RACHflow}
	\end{figure}	
	
	Due to the temporal variation of beam measurements, the access beam is selected based on measurements at time instant $ m=m_1 $ of CHO execution. This is illustrated in Figure \ref{fig:RACHflow}. Herein, the UE selects the access beam $ b_{0} \in  B^\textrm{prep}_{c_T} $ as follows
	\begin{equation}\label{eq:SelectBestPrep}
	b_{0} = \argmax_b P_{{c_T},b}^\textrm{L1}(m_1),~P_{b_{0},c_T}^\textrm{L1}(m_1) > \xi_\textrm{access},
	\end{equation}
	where $ \xi_\textrm{access} $ is the threshold that L1 RSRP beam measurements shall exceed to consider prepared beams for access. Ultimately, the UE accesses the prepared beam $ b_0 $ that satisfies the condition (\ref{eq:SelectBestPrep}) and uses the corresponding CFRA preamble. If none of the measurements $ P_{b,{c_T}}^\textrm{L1} $ of beams $ b\in B^\textrm{prep}_{c_T} $ is above the threshold $ \xi_\textrm{access} $, beam $ b_0 $ with the strongest L1 RSRP beam measurement is selected as
	\begin{equation}\label{eq:SelectStrongBeam}
	b_{0} = \argmax_b P_{{c_T},b}^\textrm{L1}(m_1).
	\end{equation}
		
	In 3GPP standardization, CBRA preambles are used if none of the L1 RSRP measurement of prepared beams is above the threshold $ \gamma_\textrm{access} $. This has the disadvantage that the UE may select CBRA resources although there are CFRA resources associated with the selected strongest beam. To tackle this issue, a Resource Efficient RACH (RE-RACH) proposed to increase the CFRA utilization as shown in green color in Figure \ref{fig:RACHflow}. Herein, the UE uses CFRA resources if the selected beam is prepared beam $ b_0\in B^\textrm{prep}_{c_T} $ even if L1 RSRP beam measurement $ P_{{c_T},b_0}^\textrm{L1} $ is below the threshold $ \xi_\textrm{access} $. This will eventually lead to less signaling and latency during the RACH procedure. UE selects either CFRA or CBRA preambles by following the process given in Fig.~\ref{fig:RACHflow} and attempts to access target cell with selected preamble. In case the access fails, UE repeats the preamble selection process and declares the HOF after several attempts. This is followed by a re-establisment process where UE searches for the new cell to be connected.
	
	\subsection{Beam-specific Enhanced Logging and Learning Approach}

	The beams $ b\in B^\textrm{prep}_{c_T} $ with the highest L3 RSRP measurements are prepared for HO as given  in (\ref{eq:PrepBeamSet}). However, RSRP of the prepared beams at $ m=m_0 $ changes over time due to temporal characteristics of the wireless channel and user mobility, e.g., RSRP of the prepared beam with highest RSRP might not be strong enough for successful random access when the access procedure is started at $m=m_1 $ and afterwards. UE is encouraged to access prepared beam with loose access threshold $ \xi_\textrm{access} $ to increase CFRA utilization which, in contrast, increases the possibility of HOF if the quality of the link between the UE and the prepared beam is not well enough for successful random access.
	
	Beam-Specific Enhanced Logging and Learning  (BELL) RACH optimization is proposed that aims to learn preparing the right beam so that the HOFs caused by wrong beam preparation of (\ref{eq:PrepBeamSet}) during the RACH process are minimized. BELL classifies the HOF events into sub-events by following a pre-defined decision tree. This is followed by an assessment mechanism that reacts on the inferences (classes), either by rewarding or penalizing the network decision on prepared beam in a way that the decisions leading to successful HO are encouraged and HOF are discouraged.

	\begin{figure}[!htb]
	\centering
	\includegraphics[width=0.8\columnwidth]{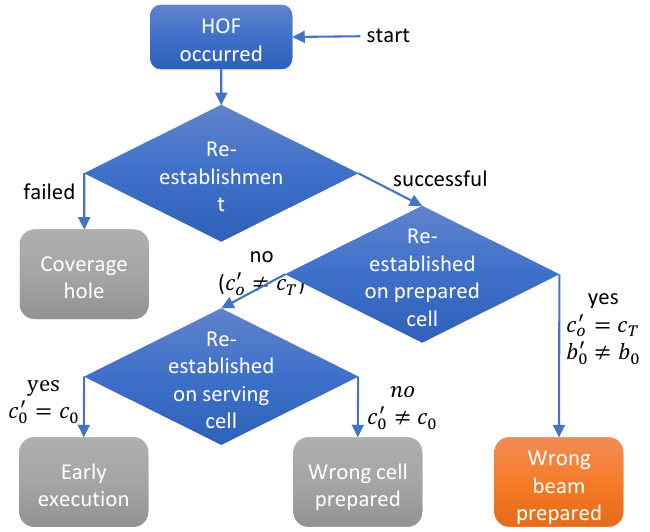}
	\caption{Classification of the HOFs with decision tree that applies the root-cause analysis on the HOFs. Classes identify the correctness of the decisions on target cell, prepared beam, and execution time of the HO process along with the problems caused by the network planning, e.g., coverage hole. Serving cell, target cell, re-established cell, prepared beam and re-established beam are denoted by $ c_0$, $c_T$, $ c_0' $, $ b_0 $ and $ b_0' $  respectively.} 
	\label{fig:decisiontree}
	\vspace{-1.2\baselineskip}
	\end{figure}
	
	Decision tree that is illustrated in Figure~\ref{fig:decisiontree} applies the root-cause analysis to classify the HOF events. HO attempt that is followed by the HOF event after several consecutive random access failures could be classified as one of 4 unique events. If the HOF is followed by failing re-establishment process, UE will be unconnected to network until further connection to a new cell. This type of HOF event is classified as "coverage hole". Although UE failed to HO to target cell $ c_T $, it may connect to $ c_0'=c_T $ after re-establishment process through a beam  $ b_0'\not\in B^\textrm{prep}_{c_T} $ other than prepared one which is interpreted as "Wrong beam prepared". This type of HOF could be avoided if HO attempt through wrong beam $ b_0 \in B^\textrm{prep}_{{c_T}} $ is avoided and preparation of  $ b_0' $ is motivated. UE may also re-establish on serving cell $ c_0 $ which shows that the HO execution condition expired earlier, before the link between the target cell $ {c_T} $ and UE became good enough for successful HO. Therefore, it is classified as "Early execution". Similarly, if UE is not connected to target cell $ c_T $ after HOF ($ c_0'\neq {c_T} $) it shows that the preparation of the cell $ {c_T} $ was not accurate. This type of HOF event is classified as "Wrong Cell Prepared". 
	
	It is already mentioned that this paper is focused on the RACH optimization by minimizing the HOF caused by beam preparation that is defined in (\ref{eq:PrepBeamSet}). Therefore, the BELL approach reacts on the HOF events those are classified as "Wrong beam prepared". To this end, a beam preparation offset $ o^\textrm{prep}_{c,b} $ is proposed and equation (\ref{eq:PrepBeamSet}) is reformulated as
	\begin{align}\label{eq:cellbeamoffset}
	B^\textrm{prep}_{c} = \{&b |P_{b,c}^\textrm{L3}(m_0) + o^\textrm{prep}_{c,b} \geq  P_{c,b_i}^\textrm{L3}(m_0)+o^\textrm{prep}_{c,b_i} , \nonumber \\
	&~b\neq b_i,~b,b_i\in B\}.
	\end{align}
	
	When HOF event is classified as "Wrong beam prepared", beam preparation offset $ o^\textrm{prep}_{c_T,b} $ of $ b\in B^\textrm{prep}_{c_T} $ is reduced $ \Delta o $ dB to penalize the preparation of this beam, and the offset $ o^\textrm{prep}_{c_T,b_0'} $ of re-established beam $ b_0'\not \in B^\textrm{prep}_{c_T} $ is increased $ \Delta o $ dB to leverage the preparation of the beam $ b_0' $ in future events. Enabling such optimization is possible only if the network logs the target cell $ c_T $ and beam $ b_0 $ information so that the root-cause analysis is applied to classify the HOF events and proper preparation decisions are learned by the network over time.
	
	Apparently, handover performance can be further optimized, e.g., HOF events that are classified as "Early execution" and "Wrong cell prepared" can be improved by conceiving the cell-specific preparation and execution offsets $ o^\textrm{prep}_{c_0,c} $ and $ o^\textrm{exe}_{c_0,c}  $, respectively, such that the early execution or wrong cell preparation can be discouraged towards specific cell. However, those classes are relatively corner cases which are rarely observed and will be discussed quantitatively in Section \ref{sec:SimRes}. RACH optimization through BELL approach defined above serves the purpose of this paper, and more extensive solutions will be presented in future study.
	
	\section{Simulation Scenario and Parameters}
	\label{sec:SimScenario}
	In this section, the investigated scenario, mobility and propagation parameters are described. These will be used to compare the different mobility performance indicators of BHO and CHO for 3GPP and RE RACH procedures, for BELL approach and for various random access beam thresholds $ \xi_\textrm{access} $.
	\begin{figure}[!htb]
		\centering
		\includegraphics[width=\columnwidth]{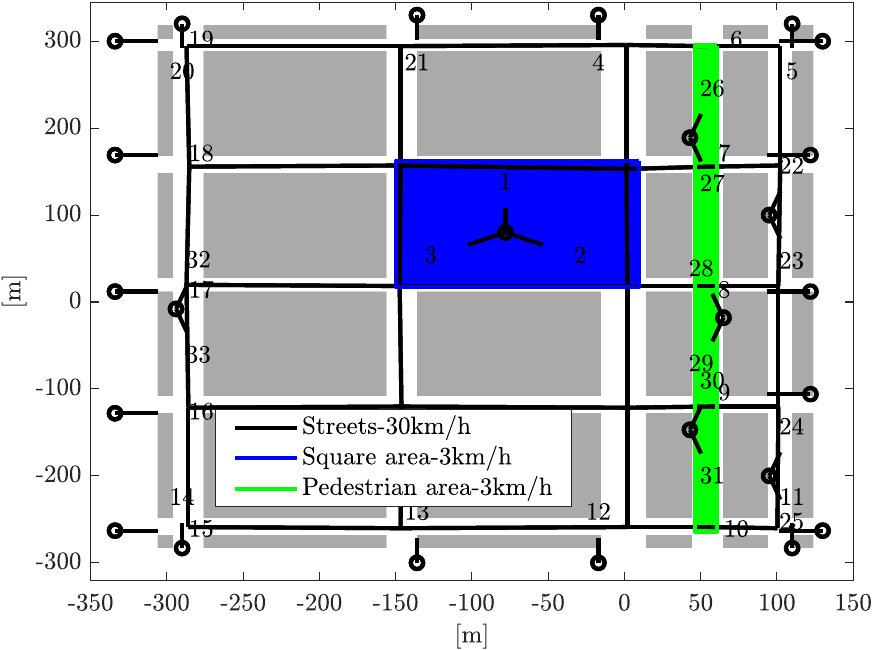}
		\caption{Madrid Grid layout is used for simulation scenario as described in METIS 2 project \cite{METIS2}. The scenario consists of buildings (grey), streets (black) with 200 users, open square (blue) with 40 users and pedestrian area (green) with 80 users.}
		\label{fig:environment}
	\end{figure}

	In this study, the Madrid Grid layout that is described in the METIS 2 project \cite{METIS2} is used. The layout is given in Figure \ref{fig:environment} and consists of buildings (grey), streets (black), open square (blue) and pedestrian area (green). There are 33 3-sector macro cells which are located on the roof tops of the buildings. The users are distributed as follows: 200 users are moving in the streets with $30$ km/h in both directions. Besides, 40 pedestrian users are walking in the open square and 80 users are walking in the pedestrian area with $3$ km/h.
	
	The scenario parameters are specified in Table~\ref{tab:sim_parameters} along with the configuration of the transmit antenna panels. Beams $ b\in[1,8]$ have smaller beamwidth and higher beamforming gain to cover far regions of the cell coverage area where beams $ b \in [9,12]$ with larger beamwidth and relatively smaller beamforming gain are defined to serve regions near to the base stations. The SINR $ \gamma_{c,b}(m) $ of a link between UE and beam $ b $ of cell $ c $ is evaluated by the approximation given in \cite{SINRModel} for the strict resource fair scheduler.
	
	\begin{table}[!htb]
		\renewcommand{\arraystretch}{1.3}
		\caption{Simulation Parameters \rom{2}}
		\label{tab:sim_parameters}
		\centering
		\begin{tabulary}{\columnwidth}{L L}
			
			\hline 
			\textbf{Parameters} & \textbf{Value}\\ 
			\hline \hline
			Carrier frequency & $28$ GHz \\ 
			
			System bandwidth & $100$ MHz \\		
				
			PRB bandwidth & $10$ MHz \\
			
			Downlink TX power & $12$ dBm/PRB \\
			
			TX antenna height & $10$ m\\
			
			TX Antenna element pattern & Table~7.3-1 in \cite{38901} \\
			
			TX panel size & $16 \times 8,~\forall b\in [1,8]$ $8 \times 4,~\forall b \in [9,12]$\\
			
			TX vertical antenna element spacing & $0.7 \lambda$  \\
			
			TX horizontal antenna element spacing & $0.5 \lambda$  \\ 
			
			Beam azimuth angle $\phi_b$ & $90,~\forall b \in [1,8]$ $97,~\forall b \in [9,12]$\\
			
			Beam elevation angle $\theta_b$ & $ -52.5+15(b-1),~\forall b\in[1,8] $\\
			& $ -45+30(b-8),~\forall b \in [9,12] $ \\
			
			Beamforming gain model & Fitting model of \cite{abstactchannel}\\
			
			RX antenna height & 1.5m\\
			
			RX antenna element pattern & isotropic \\
			
			RX antenna element gain & $0$ dBi \\
			
			Thermal noise power  & $-97$ dBm/PRB \\ 
			
			Propagation loss & deterministic model of \cite{SimplifiedDeterministic} \\
		
			Penetration loss & $0$ dB \\
			
			Fast fading model & Abstract model of \cite{abstactchannel}\\
			
			Scenario & UMi-Street Canyon \cite{38901}\\
			
			Network topology & Madrid grid \cite{METIS2} \\
			
			Number of cells & $33$\\
			
			Total number of UEs & $320$\\
			
			Number of simultaneously scheduled beams per cell & $4$\\
			
			Cell-pair specific offset $o_{c_0,c}^\textrm{A\tsub3}$ & $3$ dB\\
			
			Add offset $ o_{c_0,c}^\textrm{add} $ & $-3$ dB\\
			
			Execute offset $ o_{c_0,c}^\textrm{exe} $ & $3$ dB\\
			
			Time step size $ \Delta t $ & $10$ ms\\
			
			L1 measurement period $ \omega $ & $2$ \\
			
			Handover preparation time $ T_p $ & $20$ ms \\
			
			Number of averaged samples $ N_\textrm{L1} $ & $4$
			
		\end{tabulary}
	\vspace{-3\baselineskip} 
	\end{table}

	\textit{Handover Failure Model}: Handover failure (HOF) is a metric that is used to evaluate the mobility performance. For both 3GPP and RE RACH procedures, a UE decides to use either CBRA or CFRA preamble as shown in Figure \ref{fig:RACHflow} and attempts to access the selected beam $ b_0 $ of target cell $ c $ with the selected preamble. For successful random access, it is required that the SINR $ \gamma_{c,b_0}(m) $ of the target cell remains above the threshold $ \gamma_\textrm{out} $, during RACH procedure. A handover failure timer $ T_\textrm{304} = 500 $ ms is started when the UE starts the random access and sends the RACH preamble. The RACH procedure in Figure~\ref{fig:RACHflow} is repeated until a successful RACH attempt is achieved or $ T_\textrm{304} $ expires. In the handover failure model, a UE may succeed to access the target cell only if the $ \gamma_{c,b_0}(m) $ exceeds the threshold $ \gamma_{out} $. HOF is declared if $ T_\textrm
	{304} $ expires and the UE fails to access the target cell, i.e., $ \gamma_{c,b} < \gamma_\textrm{out} $. Once HOF is declared, the UE performs connection re-establishment which requires additional signaling and causes latency \cite{38331}.
	
	\textit{Radio Link Failure Model}: Radio link failure (RLF) is another key metric that is relevant for mobility performance. An RLF timer $T_\textrm{310} = 600 $ ms is started when SINR  $\gamma_{c_0,b}(m)$ of serving cell $ c_0 $ falls below $\gamma_\textrm{out}$ and RLF is declared if $T_\textrm{310}$ expires. During the timer, the UE may recover before detecting RLF if  SINR $ \gamma_{c_0,b} $ exceeds the second threshold $\gamma_\textrm{in}$ which is higher than $\gamma_\textrm{out}$. A detailed explanation of the procedure is given in \cite{38331}.
	
	\section{Performance Evaluation}
	\label{sec:PerformanceEval}
	In this section, mobility performance of RE RACH is compared against that of 3GPP RACH for both BHO and CHO. Then, the performance analysis is extended for BELL approach. The key performance indicators (KPIs) used for comparison are explained below. 
	
	\subsection{KPIs}
	\label{sec:KPIs}
	
	\subsubsection{CBRA Ratio (R\tsub{CBRA})} Total numbers of successful CBRA and CFRA procedures that are observed during a mobility simulation are denoted by $ N_\textrm{CBRA} $ and $ N_\textrm{CFRA} $, respectively. The fraction of CBRA events in a simulation is formulated as
	\begin{equation}\label{CBRA rate}
	R_\textrm{CBRA}[\%] = \frac{N_\textrm{CBRA}}{N_\textrm{CBRA}+N_\textrm{CFRA}}\times 100\%.
	\end{equation}
	
	\subsubsection{$ N $\tsub{HOF}} Total number of HOFs that are observed during a simulation. 
	
	\subsubsection{$ N $\tsub{RLF}} Total number of RLFs that are declared in the network.
	
	Both $ N_\textrm{HOF} $ and $ N_\textrm{RLF} $ are normalized to number of UEs and simulation time as illustrated in the following section.
	\subsection{Simulation Results}
	\label{sec:SimRes}
	
\begin{figure}[!htb]
		\centering
		\includegraphics[width=\columnwidth]{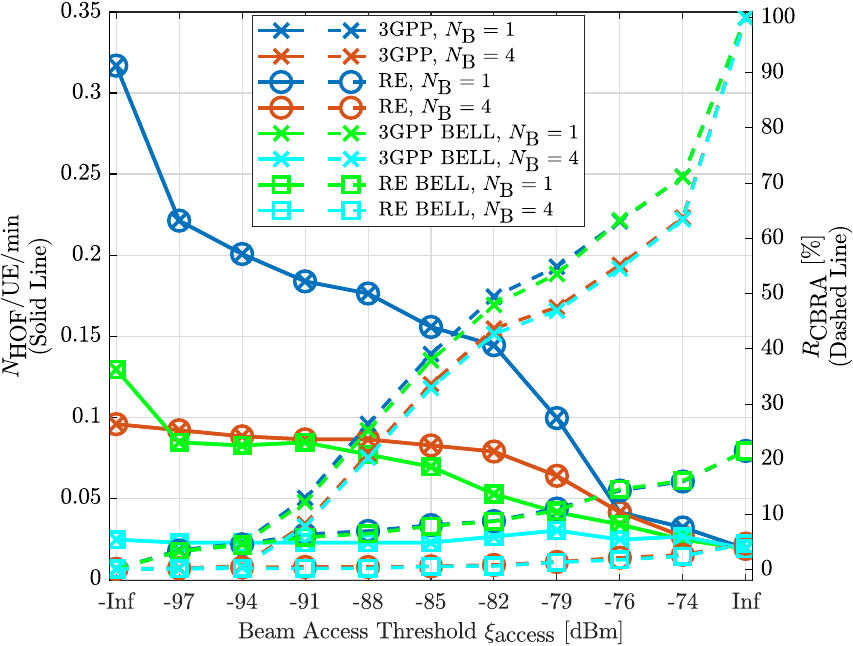}
		\caption{The number of HOFs and ratio $ R_\textrm{CBRA} $ are shown for CHO as a function of beam access threshold $ \xi_\textrm{access} $. The evaluation is given for 3GPP and RE RACH procedure and for different number $ N_\textrm{B} $ of beams as parameters. The performance of the BELL approach is also revealed for both RACH procedures.}
		\label{fig:CHO}
		\vspace{-1.5\baselineskip}
	\end{figure}

	The mobility performances of the 3GPP and RE RACH procedures and BELL approach is investigated for both CHO and BHO, and for the given scenario in Figure \ref{fig:environment}. To this end, the impact of different beam access thresholds $ \xi_\textrm{access} $ values and number of prepared beams $ N_\textrm{B} = \vert B_\textrm{prep,c}\vert$ on the aforementioned mobility KPIs of Section~\ref{sec:KPIs} are analyzed. Figure~\ref{fig:CHO} and Figure~\ref{fig:BHO} show the number $ N_\textrm{HOF} $ of handover failures per UE$\cdot$minutes (UE$\cdot$min) with solid line on the left axis and CBRA ratio $ R_\textrm{CBRA} $ with dashed line on the right axis as a function of $ \xi_\textrm{access} $ (in dB) for CHO and BHO, respectively. The results of 3GPP and RE RACH procedures are shown for different number of prepared beams $ N_\textrm{B}=1 $ and $ N_\textrm{B}=4 $. Performance of BELL approach on CHO is also illustrated in Figure~\ref{fig:CHO} for different RACH procedures and $ N_\textrm{B} $.
	
	\subsubsection{3GPP and RE RACH Performance on CHO}
	\label{sec:3GPP-RE-RACH-CHO}
	Figure~\ref{fig:CHO} shows that for $ \xi_\textrm{access} = -\infty$ the UE uses only CFRA preambles ($ R_\textrm{CBRA}=0$) for all RACH procedures since the UE always selects a prepared beam from set of $ B^\textrm{prep}_{c_T} $ of target cell $ c_T $. On the other hand, $ \xi_\textrm{access} = -\infty$ leads to worst HOF performance because the received signal power of the prepared beam changes over time and the prepared beam does not always remain a good candidate during the time between handover preparation and execution phases. Ultimately, the SINR $ \gamma_{c,b_0}(m) $, $ m\geq m_1 $ of the accessed beam $ b_0 $ falls below $ \gamma_\textrm{out} $ which leads to HOF. This is more visible for $ N_\textrm{B}=1 $ since the UE does not have any other options for selecting another prepared beam. Increasing $ N_\textrm{B} $ from $ 1 $ to $ 4 $ reduces the access failure $ N_\textrm{HO} $ to one third of its value since it increases the chance of selecting the strongest beam.

	For increasing values of access threshold $ \xi_\textrm{access} $, the RACH beam selection procedure prioritizes the L1 RSRP beam measurements $ P_{c,b}^\textrm{L1}(m) $ and the UE becomes less persistent on selecting one of the prepared beams. As a consequence, beams with higher $ P_{c,b}^\textrm{L1}(m) $ are selected to be accessed which yields higher $ \gamma_{c_0,b}(m)$ and less HOFs. On the other hand, for higher $ \xi_\textrm{access} $, UE tends to select prepared beams less frequently which results in more likely use of CBRA preambles for random access. However, it is observed that the ratio of CBRA resource usage is much smaller for the RE RACH procedure for higher $ \xi_\textrm{access} $. This is because the UE still use CFRA resources if none of the prepared beams have beam measurements above threshold $ \xi_\textrm{access} $.
	
	Results in Figure~\ref{fig:CHO} also show that the number of HOFs of the 3GPP and RE RACH procedures reaches their lowest value at $ \xi_\textrm{access}=\infty $ and is the same for both $N_\textrm{B}=1$ and $N_\textrm{B}=4$. This is because the beam of the target cell with the strongest L1 RSRP measurement is selected in both RACH procedures regardless of the set of prepared beams $ B^\textrm{prep}_c $. Hence, the selected beam $ b_0 $ of target cell $ c $ with strongest measurement $ P_{c,b_0}^\textrm{L1}(m)$ leads to higher SINR $ \gamma_{c,b_0}(m) $ and in turn lower HOF. However, CBRA ratios of the 3GPP and the RE RACH procedures diverge significantly at $ \xi_\textrm{access}=\infty $. In particular, for the 3GPP RACH procedure, the UE selects only CBRA preambles for random access for any $ N_\textrm{B} $ value since all prepared beams have L1 measurements that are below $ \xi_\textrm{access} $. This is not the case for the RE RACH procedure because preamble selection still considers the prepared beams although that L1 measurement is not above $ \xi_\textrm{access} $.
	
	Furthermore, the same HOF performance of CHO is observed for both the 3GPP and the RE RACH procedures since the HOF depends on the selected beam and both RACH procedures do not differ with respect to beam selection procedure as shown in Figure~\ref{fig:RACHflow}.
	
	\subsubsection{BELL Performance on CHO}
	BELL performance on CHO is given in Figure~\ref{fig:CHO} for 3GPP and RE RACH procedures and for different number $ N_\textrm{B} $ of prepared beams. HOF rate of RACH procedures without BELL approach is approximately $ 0.3 $/UE/min for $ \xi_\textrm{access}=-\infty $ and for $ N_\textrm{B}=1 $ (only CFRA). Applying BELL approach on both RACH procedures prevents around $ 60\% $ the HOF and reduces it to $ \sim 0.1 $/UE/min. For $ \xi_\textrm{access} > -\infty $, BELL approach with $ N_\textrm{B} = 1$ provides less HOF, even when it is compared to the performance of 3GPP and RE RACH procedures without BELL approach and $ N_\textrm{B}=4 $, and the results coincide at $ \xi_\textrm{access}=\infty$ as explained in Section \ref{sec:3GPP-RE-RACH-CHO}. This shows that the BELL reveals a better results by not only increasing the HOF performance but also requiring less resources to be reserved (smaller $ N_\textrm{B} $). 
	
	When the number $ N_\textrm{B} $ of prepared beams is increased from $ 1 $ to $ 4 $, HOF performance of BELL approach remains constant for all $ \xi_\textrm{access} $ values and shows the same performance of $ \xi_\textrm{access} = \infty$ case where the strongest beam is selected as access beam, no matter which beam was prepared. Considering the fact that BELL approach does not react on all classes that are defined for HOFs in Figure~\ref{fig:decisiontree} (e.g., coverage hole, early execution and wrong cell preparation), 
	
	Resudial HOFs that are observed for $ \xi_\textrm{access} =\infty$ reveals the HOFs that are caused by the classes that BELL approach does not react, i.e., "coverage hole", "early, execution", and "wrong cell preparation" classes that are defined for HOFs in Figure~\ref{fig:decisiontree} are not covered in BELL approach. Considering the fact that $ \xi_\textrm{access} =\infty$ shows the best HOF performance of beam preparation, BELL approach of $ N_\textrm{B}=4 $ case provides an optimum operation point at $ \xi_\textrm{access} =-\infty$ since it gives full CFRA resource utilization ($ R_\textrm{CBRA}=0\% $) for minimum achievable HOF. 
	
	CBRA rate $ R_\textrm{CBRA} $ is slightly reduced when the BELL approach is employed. This is because BELL approach designates the prepared beam based on a rewarding algorithm which leverages the beams with higher L1 RSRP beam measurements at the time of access $ m>m_1 $ and the chance of preparing a beam is indirectly increased where the prepared beams more likely to satisfy the CFRA condition.

	\subsubsection{RACH Performance Comparison between BHO and CHO}
\begin{figure}[!htb]
		\centering
		\includegraphics[width=\columnwidth]{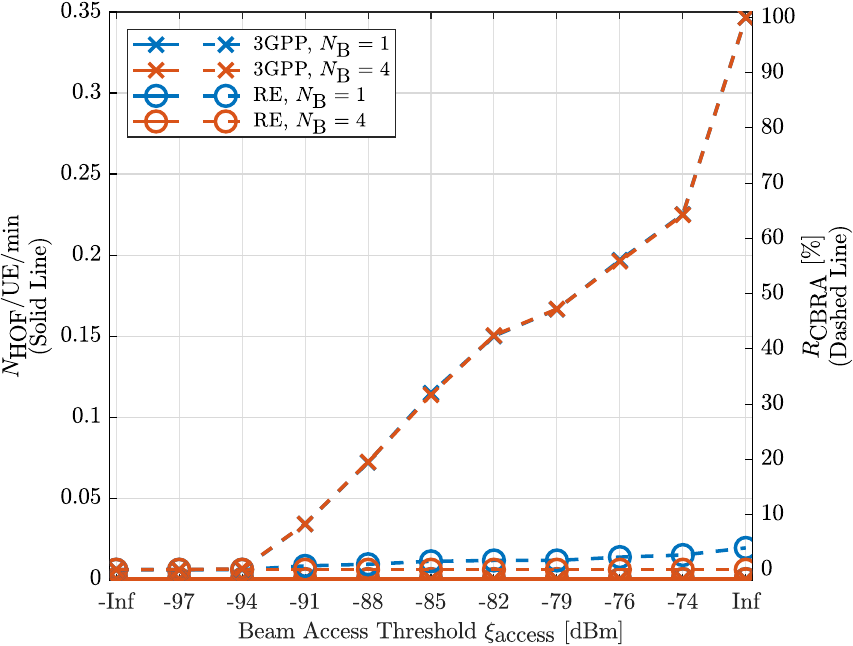}
		\caption{The number of HOFs and ratio $ R_\textrm{CBRA} $ are shown for BHO as a function of beam access threshold $ \xi_\textrm{access} $ with RACH procedure and number $ N_\textrm{B} $ of beams as parameters.}
		\label{fig:BHO}
		\vspace{-1.4\baselineskip}
	\end{figure}

	Figure~\ref{fig:BHO} shows that HOF is not observed at BHO for any number $ N_\textrm{B} $ of prepared beam and beam access threshold $ \xi_\textrm{access} $. This is because, compared to the CHO results in Figure~\ref{fig:CHO}, the time $ T_\textrm{p} $ that elapses between preparation and the phases of BHO is shorter than that of CHO ($T_p + T_0 $) and during this time the measurements of the prepared beams do not change. Consequently, UEs performs access to a beam $ b_0 $ that yields sufficient $ \gamma_{c,b_0}(m) $ at target cell $ c $. BELL approach cannot improve the HOF performance of BHO since the BELL aims to reduce the HOFs and the HOF is not observed in BHO.
	
	Figure~\ref{fig:BHO} also shows that the CBRA ratio of the proposed RACH procedure slightly increases for higher $ \xi_\textrm{access} $ because the measurements of the beams do not change much between preparation and access phases which is shorter than that of the CHO case. However, the CBRA ratio of the 3GPP procedure in Figure~\ref{fig:BHO} gradually increases for increasing $ \xi_\textrm{access} $ as it is observed for the CHO case in Figure~\ref{fig:CHO}. This is also due to the fact that the 3GPP RACH procedure does not consider the prepared beams in case the L1 measurements are below the access threshold $ \xi_\textrm{access} $.

	\subsubsection{Failure Results}
\begin{figure}[!htb]
		\centering
		\includegraphics[width=0.9\columnwidth]{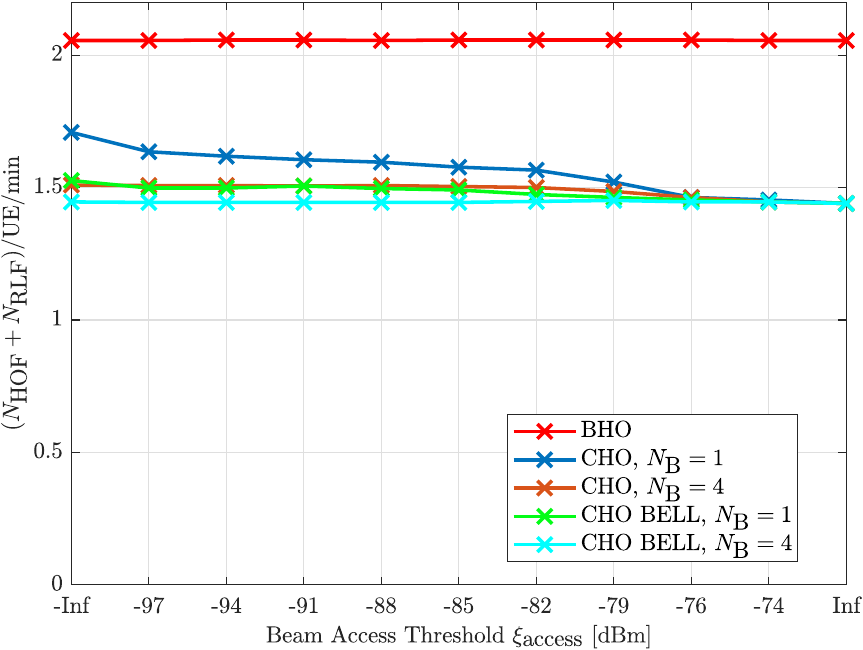}
		\caption{The total number of failures is shown for CHO and BHO case as a function of beam access threshold $ \xi_\textrm{access} $ with number $ N_\textrm{B} $ of beams as parameter.}
		\label{fig:RLF}
	\end{figure}

	Figure~\ref{fig:RLF} shows the total number of failures $ N_\textrm{HOF} + N_\textrm{RLF} $ per UE$\cdot$min as a function of the beam access threshold $ \xi_\textrm{access} $ for both CHO and BHO and for BELL approach. As it has been shown in Figure~\ref{fig:CHO} and \ref{fig:BHO} that failure rate is independent of the RACH procedure, the results in Figure~\ref{fig:RLF} do not differentiate the two RACH procedures. Besides, the total number of failures for BHO shows the same performance for both $ N_\textrm{B}=1 $ and $ N_\textrm{B}=4 $ and for BELL approach.
	
	Although Figure~\ref{fig:BHO} shows that the mobility performance of BHO on HOF is better than that of CHO, Figure~\ref{fig:RLF} illustrates that the overall failure performance of BHO is improved by the conditional execution mechanism that is introduced by CHO. Furthermore, one can also state that the failures that are observed in the mobility scenario are dominated by RLF and this is improved by CHO despite the HOF increase that is observed for CHO compared to BHO, see Figure~\ref{fig:CHO} and Figure~\ref{fig:BHO}. The enhancements of the larger $ N_\textrm{B} $ and BELL approach on overall performance of CHO is also visible in Figure~\ref{fig:RLF}.

	\section{Conclusion}
	\label{sec:Concl}
	
	In this paper, conditional handover of 3GPP release 16 is analyzed for NR beamformed systems. Baseline and conditional handover procedures have been reviewed along with L1 and L3 UE measurements that are relevant for mobility. In addition, the 3GPP random access procedure is revisited and a new random access procedure is proposed that aims to increase contention-free random access and reduce in turn signaling overhead and latency during handover. Besides, decision tree based supervised learning method is proposed to reduce the HOF caused by beam preparation step of RACH procedure. The results show that the optimum operation point is achieved with the proposed learning algorithm. Furthermore, the mobility performance of conditional handover is compared against baseline handover. Simulation results have shown that the number of fall-backs to contention based random access is reduced significantly when the proposed random access procedure is used.
		
	Moreover, the results have revealed that the baseline handover procedure causes less handover failures than  conditional handover. However, the total number of failures for conditional handover is less than that of baseline handover due to the de-coupled handover preparation and execution phases, providing mobility robustness. 
		
	\bibliographystyle{IEEEtran}
	\bibliography{references}

\end{document}